\documentclass[aps, prd, amsmath, floats, floatfix, twocolumn, nofootinbib]{revtex4-2}
\usepackage{graphicx}
\usepackage{color}
\usepackage{latexsym}
\usepackage{amsthm,amsmath,amssymb}
\usepackage{mathrsfs}
\usepackage{amssymb}

\newcommand{\be}{\begin{eqnarray}}
\newcommand{\ee}{\end{eqnarray}}

\newcommand{\bea}{\begin{eqnarray}}
\newcommand{\eea}{\end{eqnarray}}

\begin{document}

\title{Geodesic incompleteness of some popular regular black holes} 

%\author{Fabio Briscese}\email{briscese.phys@gmail.com, briscesef@sustc.edu.com}

\author{Tian Zhou}
\email{11930538@mail.sustech.edu.cn}

\author{Leonardo Modesto}
\email{lmodesto@sustech.edu.cn}

\affiliation{Department of Physics, Southern University of Science
and Technology, Shenzhen 518055, China}

\begin{abstract}

Throughout the study of the geodesics of some popular spherically symmetric regular black holes, we hereby prove that the analytically extended Hayward black hole is geodetically {\em incomplete}. The simplest extension of the Culetu-Simpson-Visser's non-analytic smooth black hole is also geodetically {\em incomplete}, with the exception of the antipodal continuation of the radial geodesics. However, the huge ambiguity in the extension of non analytic spacetimes is tantamount of geodesic incompleteness and such spacetimes do not solve the singularity issue unless at least all the extensions turn out to be complete. 
%However, even for the latter spacetime the rotating spacetime. 
Hence, we provide several mere modifications of such spacetimes in order to make them geodetically complete in all possible extensions beyond $r=0$.

\end{abstract}

\maketitle

\section{Introduction}
In order to tame the black holes' singularity problem in Einstein's gravity, many regular spacetimes have been proposed regardless of an underlying dynamic \cite{Bardeen, Hayward:2005gi, Bonanno:2000ep, Modesto:2009ve, Nicolini:2005vd, Modesto:2010uh, Modesto:2010uh, Burzilla:2020utr, Giacchini:2018wlf, Modesto:2008im,Culetu:2013fsa, Culetu:2014lca,Xiang:2013sza,Simpson:2019mud}, namely they are just formal solutions of the Einstein's equations for a proper choice of the energy-momentum tensor that violate some or all the energy conditions. So far it turns out that the Kretschmann scalar $R_{\mu\nu\rho\sigma}R^{\mu\nu\rho\sigma}$ and other curvature invariants are regular at the center of the proposed regular black holes. However, the geodesic completion has not been studied properly and/or completely and will be the main topic of this paper. 

For the sake of simplicity, let us consider a general spherically symmetric solution in Schwarzschild coordinates, namely\footnote{We here work in Planck units, namely the Newton's gravitational constant, the speed of the light, and the Plank's constant are: $c= G=\hbar =1$.}
\be
ds^2 = - \left( 1 - \frac{2 M(r)}{r} \right)dt^2 + \frac{dr^2}{1 - \frac{2 M(r)}{r}} + r^2 d \Omega^{(2)} \, , 
\label{SchLINE}
\ee
where the mass is now allowed to be a function of the radial coordinate. When $M(r)$ is constant we are back to the Schwarzschild solution. 

The spacetimes we will consider in this paper have regular Kretschmann scalar $R_{\mu\nu\rho\sigma}R^{\mu\nu\rho\sigma}$ for $r\geqslant 0$, although for some metrics the Kretschmann invariant can still be divergent somewhere for $r<0$. However, the curvature invariants do not have a real physical meaning and only provide a tool for a simple and preliminary analysis of spacetime's regularity. In physics, we have to look at the geodesic equations and see whether and where they are well defined. We here remind the case of the Taub-NUT's metric that has finite Kretschmann invariant, but it is not geodetically complete. Therefore, the very basic requirement for regular spacetimes should be the completion of all causal geodesics. 

For the sake of simplicity, let us here focus on massive probe particles moving in the spacetime (\ref{SchLINE}). The radial geodesics satisfy the following equation in proper time, 
\be
\dot{x}^2 = -1 \quad \Longrightarrow \quad g_{tt}\dot{t}^2+g_{rr}\dot{r}^2=-1\, ,
\label{xdot}
\ee
where $-g_{tt}=g_{rr}^{-1}=1-2M(r)/r$ and the dot represents the derivative respect to the proper time $\tau$. Replacing the conserved quantity corresponding to the time translation invariance, namely $e\equiv -g_{tt}\dot{t}$, in (\ref{xdot}), (where $e$ is a dimensionless constant) we end up with the following equation for the radial motion, 
\be
\dot{r}^2 = e^2 - 1 + \frac{2 M(r)}{r}\, . 
\label{radial}
\ee
Choosing now as initial radial position $r_i$, the proper-time for a massive particle to move from $r_i$ to $r < r_i$
is given by
\be\label{propertime}
\tau(r)=\int_r^{r_i}dr \frac{1}{\sqrt{\dot{r}^2}} \, .
\ee
Since in (\ref{radial}) $\dot{r}^2 > 0$, the right hand side of the above radial geodesic equation (\ref{radial}) is only defined for values of $r$ compatible with:
\be
e^2 - 1 + \frac{2 M(r)}{r} \geqslant 0 \, . 
\ee
If the metric we are considering is regular in $r = 0$ (most of the known regular spacetimes have a de Sitter's core in $r=0$), we also have to check if the proper time $\tau$ to reach such point is finite or infinite. 

If $\tau$ is finite (like in the de-Sitter core case), then, we are forced to extend the spacetime beyond $r=0$ to negative values of the radial coordinate $r$. There is nothing strange in defining the radial coordinate to be negative because in general relativity the coordinates have no physical meaning. Now for negative values of the radial coordinate two remarkable things may happen: (i) there exists at least a point $r_0$ in which 
$\dot{r}^2 \rightarrow \infty$, namely there is at least one pole, or (ii) $\dot{r}^2$ is always finite $\forall \, r$. In the latter case, $\dot{r}^2$ can change sign depending on the value of the energy $e$. Indeed, for $e$ small enough $\dot{r}^2$ can be zero let's say in the point $r_0$. 
On the other hand if $e$ is large enough $\dot{r}^2 > 0$ $\forall \, r$. %
Let us expand on points (i) and (ii). 

In the first case, near the point $r_0$, we have:
\be 
\dot{r}^2 \sim (r - r_0)^{-n} \, , \quad n>0\, ,
\ee
where the power $n$ can be a real number. Hence, the proper time to reach the point $r_0$ is finite, but 
%as pointed out above 
the geodesic equation (\ref{radial}) is singular $\forall \, e$ and we can not extend the geodesic motion beyond $r_0$. Moreover, the causal structure of the spacetime does not allow the particle to bounce back because near $r_0$ the radial coordinate is time-like, exactly like for the Schwarzschild metric for $r\gtrsim0$.

In the second case, close to $r_0$ (for $e$ small enough), we usually have: 
\be
\dot{r}^2 \sim (r - r_0)\, .
\label{r0p}
\ee
Again, according to (\ref{propertime}), the proper time to reach $r_0$ is finite. Remarkably, 
if (\ref{r0p}) holds $\dot{r}^2$ changes sign in $r_0$, which implies that massive particles will bounce back because of the potential barrier whether their energy is not big enough. On the other hand for $e$ large enough $\dot{r}^2 > 0$ $\forall \, r$. 
It deserves to be noticed that for some probe particles with a special value of the energy, $\dot{r}^2 \sim (r - r_0)^n$, where the power $n$ is an integer such that $n\ge 2$. These massive particles take an infinite amount of proper time to reach $r_0$.

Looking at the effective potential 
\be 
V(r) = - g_{tt}(r) = 1- \frac{2M(r)}{r} \, ,
\ee
if it is finite in the whole spacetime, the particles with energy large enough can travel to arbitrarily negative values of the radial coordinate $r$ therefore the spacetime is geodetically complete. 
On the other hand, if the effective potential tends to positive infinity at a certain point $r_0$, every massive particle will bounce back. However, massless particles may arrive at $r_0$ for a finite value of the affine parameter. Indeed, the null geodesics satisfy the equation $\dot{r}^2=e^2$ (this is what actually happens for the Reissner-Nordström black hole in $r=0$). Looking at the curvature invariants, the Kretschmann scalar will diverge where $g_{tt}$ is singular and there is a curvature singularity in $r=r_0$. Therefore, it is easy to show that the spacetime is complete for time-like geodesics, but incomplete for null geodesics.

We conclude that the geodesic completion of the metric (\ref{SchLINE}) is tantamount the absence of singularities in the $g_{tt}(r)$ component of the metric $\forall \, r\in[-\infty,+\infty]$.

The analysis of above focuses on the analytic properties of the function $M(r)$ for all values of the radial coordinate $r\in\mathbb{R}$. However, if for $r\rightarrow 0$ the function $M(r)/r\rightarrow 0$ at least quadratically, then the metric can be simply truncated to positive values of $r$ and the geodesic motion extended again to positive values of $r$. In other words, a probe particle falling down towards $r = 0$ (with decreasing value of $r$ and fixed angle $\phi_0$ in the equatorial plane) will bounce back at the angle $\phi_0+\pi$: {\em antipodal extension}.

Notice that although the metrics we are going to consider in this paper approach the Minkowski spacetime for $r\rightarrow0$, they are very different from the latter one. Indeed, the particular coordinate transformation $r\rightarrow -r$ is an isometry for the Minkowski spacetime, but it is not such for the black holes that we will consider in this paper. Therefore, unlike the Minkowski spacetime, the continuation of black holes is ambiguous and all possible extensions must be considered. Indeed, since we do not have a trustable tool (like fundamental equations of motion) for selecting one out of all possible extensions, we can assert that \emph{a metric is genetically complete only if all the extensions are}.

For the particular case of the nonanalytic spacetimes, such as the Culetu-Simpson-Visser \cite{Simpson:2019mud}, 
the huge ambiguity in the extension of the metric beyond $r=0$ is actually equivalent to the geodesic incompletion. Indeed, in physics, in physics there is no difference between not been able to predict the future or to have to chose between a large of even infinite number of possible future configurations.

For analytic metrics, such as the Hayward spacetime and generalizations (see later in the paper), if we select out the antipodal extension as the only doable physical one, it turns out that such antipodal continuation is not analytic for some regular black holes (an exhaustive discussion is provided in the appendix for the case of the Hayward metric). In general relativity, the analyticity of the continuation is usually required in order to preserve the uniqueness of continuation and predictability. 

Finally, we make a comment about the natural quantum selection of the physical spacetimes. Some general results are in favor of analytic metrics obtained in the framework of the finite action principle \cite{nature} in quantum gravity. According to such analysis, non-smooth metrics are filtered out by the path integral if certain higher-derivative terms are included in the gravitational action. \cite{Giacchini:2021pmr}.
%Moreover, due to the higher-derivative invariants in the gravitational effective action, unsmooth metrics will be filtered out by the action principle selection (see \cite{Giacchini:2021pmr}). So it is safe to assume that metrics (or geodesics) are at least $C^\infty$ continuous. 

%it is not sure that there is no observable physical quantity relating to higher derivatives of metrics. 

%{\color{red}
%Finally, another remarkable reason to extend the spacetime to negative values of the radial coordinate $r$ is that the simple antipodal continuation of geodesics is usually not analytic in the case of rotating black holes. For example, it is well known that the Kerr spacetime must be extended to the region of negative $r$ even in General Relativity. %Likewise, if the metric (\ref{SchLINE}) is perturbed by adding a weak spin, the spacetime should be extended to the region with negative coordinate $r$. 
%Therefore, for rotating regular black holes (see \cite{Bambi:2013ufa} for a recipe about how to get regular rotating black holes starting from spherically symmetric black holes), in spite of the disappearance of the singularity in $r=0$, it is necessary to check whether there is any other singularity for $r<0$. Hence, our analysis for spherically symmetric black holes is crucial in order to infer whether the corresponding rotating black holes are indeed singularity free. 
%}

In this paper, we will focus on the geodesic completeness of two very popular regular black holes and point out that they are actually geodetically incomplete. 

%%%%%%%%%%%%%%%%%%%%%%%%%%%%%%%%%%%%%

\section{Geodesic Incompleteness of Hayward's black hole}

We here consider the spherically symmetric and static Hayward's metric \cite{Hayward:2005gi}, namely 
\be 
ds^2=-f(r)dt^2+f^{-1}(r)dr^2+r^2d\Omega^{(2)}\, ,
\label{Hayward0}
\ee
where the function $f(r)$ takes the form
\be
\label{heywardf}
f(r)=1-\frac{2Mr^2}{r^3+L^3}\, ,
\ee
in which $M$ is a constant mass and $L$ is a length scale. %The metric is the so-called Hayward black 
%$L$ is a small constant length. 
The metric asymptotically approaches the Schwarzschild spacetime, but near $r=0$ the metric shows a de Sitter core because $f(r)\sim1- (2M/L^3) r^2$, contrary to the Schwarzschild singularity present at the center of the black hole. 

Let us now focus on the issue of the geodesic incompleteness of the Hayward black hole. Consider a probe massive particle that falls radially into the black hole. The radial geodesic equation of motion of a test particle is given by
\be
\dot{r}^2=e^2-1+\frac{2Mr^2}{r^3+L^3},\quad e\equiv -g_{tt}\dot{t} \, ,
\ee
while the effective potential reads:
\be
V\equiv 1-\frac{2Mr^2}{r^3+L^3} \, .
\ee
\begin{figure}
\begin{center}
\hspace{-0.2cm}
\includegraphics[height=4cm]{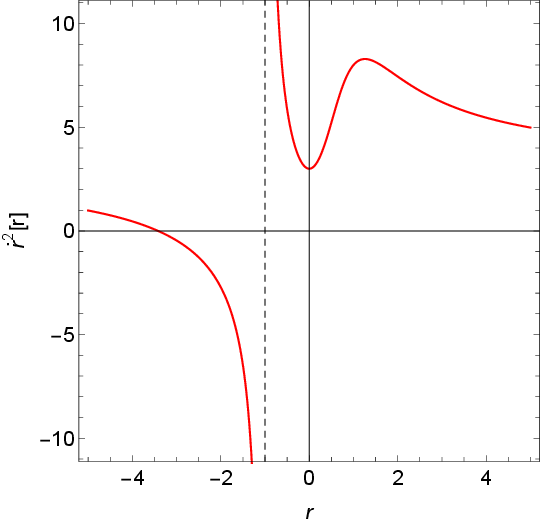}
\hspace{0.1cm}
\includegraphics[height=4cm]{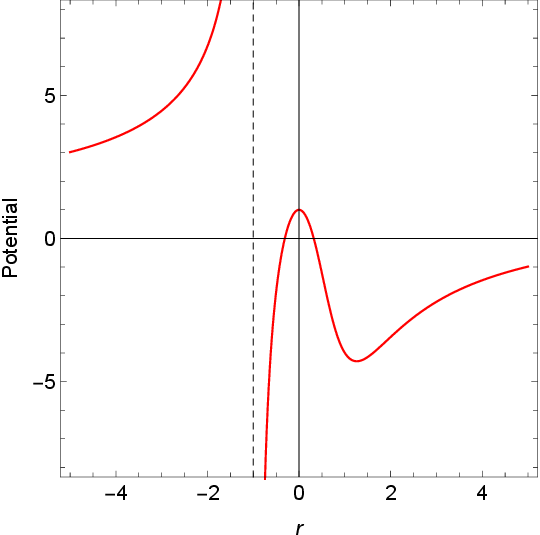}
\end{center}
\caption{The left plot shows $\dot{r}^2(r)$ for $e=2$, $M=5$ and $L=1$. The right plot shows the effective potential $V= - g_{tt}(r)$ for the Hayward spacetime with $M=5$ and $L=1$.}
\label{pic1}
\end{figure}

The plots of $\dot{r}^2$ and $V$ are shown in Fig. \ref{pic1}. A probe particle with energy $e < 1$ can not overcome the potential barrier and will bounce back consistently with the causality structure of the spacetime. Indeed, $g_{tt} < 0$ for $r \sim 0$. On the the other hand, for $e > 1$, $\dot{r}^2 \sim e^2-1$ when $r\rightarrow0$. Integrating equation (\ref{propertime}) for the case of the Hayward metric, we find that a probe particle can reach the point $r=0$ in a finite amount of proper time. Moreover, in the Appendix we have shown that the geodesics should be analytically extended to the region with negative values of $r$ because the antipodal extension is not analytic. Therefore, massive particles can reach the region $ - L \leqslant r \leqslant 0$ in finite proper time, and the point $r = - L$ turns out to be a singularity of the Hayward metric. Indeed, the spacetime structure of the Hayward metric near $r = - L$ is very similar to the Schwarzschild one near $r=0$. In particular, $\dot{r}\sim (r+L)^{-1}$ for $r\rightarrow - L$, and after integrating (\ref{propertime}) in the radial coordinate $r$ starting from a general initial position, the proper time to reach $r = - L$ turns out to be finite. Therefore, the geodesics will end in $-L$ in finite proper time and the extended Hayward spacetime is geodetically incomplete.

%We here implemented the unique spacetime continuation to negative values of $r$ consistently with the request that $f(r)$ is a $C^\infty$ function. 

% with we used here is based on the function $f(r)$, which is analytic for negative values of $r$. Indeed the Hayward metric is unique whether we require the components of the metric to be $C^\infty$ continuous condition.

%%%%%%%%%%%%%%%%%%%%%
\section{Generalized Hayward metric}

We hereby consider the following generalization of the Hayward metric, %which the function $f(r)$ takes the form
\be
\label{f}
f(r)=1-\frac{2Mr^{n-1}}{r^n+L^n}, \, \quad n \in \mathbb{N} \,, \quad n\ge3. 
\ee
%{\color{red} in which the power $n$ is an integer}. 
The metric defined in terms of the function (\ref{f}) is again asymptotically Schwarzschild, but near $r=0$, $f(r)\sim 1- (2M/L^n) r^{n-1}$, and the Kretschmann scalar $R_{\mu\nu\rho\sigma}R^{\mu\nu\rho\sigma}\sim r^{2n-6}$ turns out to be regular when $n\ge 3$. Therefore, for the generalized Hayward metric, 
there is no curvature singularity in $r\in [0,+\infty)$. 

\begin{figure}
\begin{center}
\includegraphics[height=2.8cm]{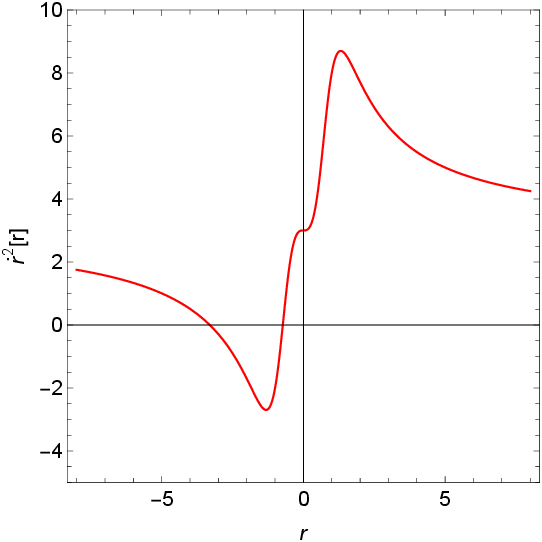}
\includegraphics[height=2.8cm]{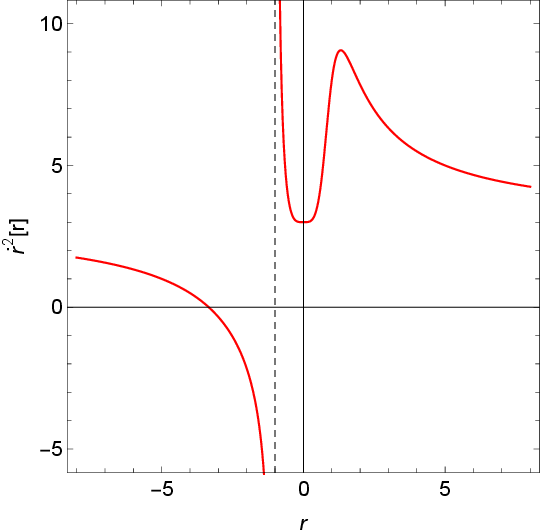}
\includegraphics[height=2.8cm]{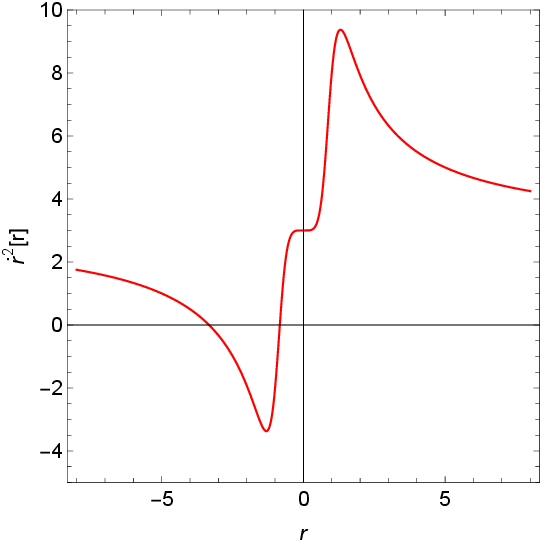}
\end{center}
\caption{$\dot{r}^2(r)$ for $e=2$, $M=5$, and $L=1$. From the left to the right, the values of $n$ are $4$, $5$, and $6$ respectively.}
\label{r2}
\end{figure}

Focusing on the geodesic completeness of the metric (\ref{Hayward0}) with (\ref{f}), the equation of motion for radial time-like geodesics reads:
\be\label{eom}
\dot{r}^2=e^2-1+\frac{2Mr^{n-1}}{r^n+L^n}\, .
\ee
For massive particles with $e^2>1$, the proper time to reach $r=0$ is finite. Hence, as a general conclusion of the Appendix, we need to extend the coordinate $r$ to negative values (a plot of $\dot{r}^2$ as a function of $r$ is given in Fig.\ref{r2}).

If $n$ is even, $\dot{r}^2$ is regular everywhere. However, where $\dot{r}^2$ is negative the probe particle does not have enough energy to overcome the potential barrier and it bounces back. On the other hand, for probe particles with energy large enough $\dot{r}^2$ is always positive, and the geodesics can be extended to $r = - \infty$. Since the metric (\ref{f}) is asymptotically flat, the proper time to reach $r= -\infty$ is infinite. One can also show that the affine parameter to approach $r = -\infty$ diverges for massless particles. Hence, the spacetime with (\ref{f}) is complete for massive as well as massless particles.

If $n$ is odd, $\dot{r}^2$ has a pole at $r = - L$, and close to the pole, the spacetime structure is similar to Schwarzschild's spacetime in $r=0$. Thus massive particles can reach the pole in a finite amount of proper time. Since these geodesics end in $r = - L$ the spacetimes with $n$ odd are geodetically incomplete.

\begin{figure}
\begin{center}
\includegraphics[height=4.5cm]{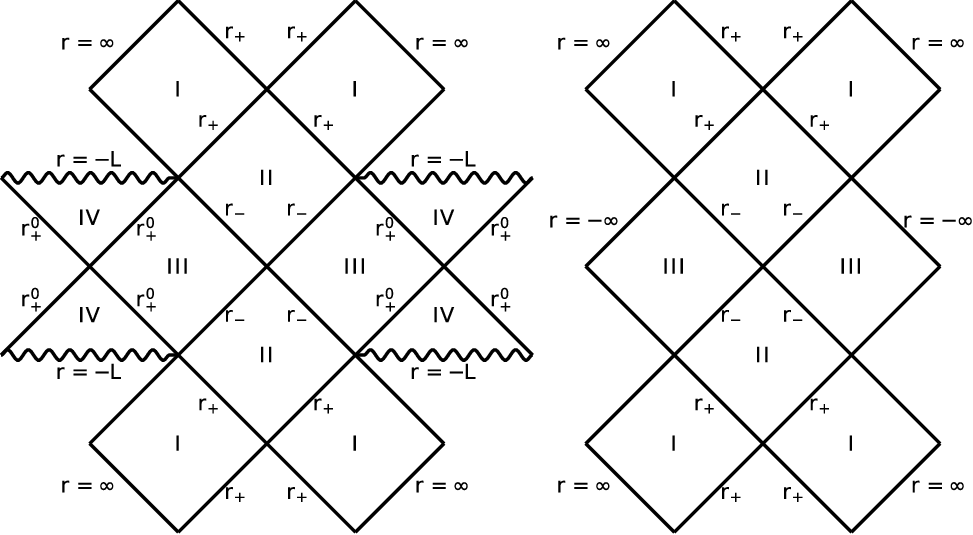}
\end{center}
\caption{Panel on the left: maximal extension of generalized Hayward's spacetime when $n$ is an odd positive integer. It deserves to be noticed that the diagram is also infinitely extendible in both the left and right directions.
Panel on the right: maximal extension of generalized Hayward's spacetime when $n$ is an even positive integer.
}
\label{pd}
\end{figure}

To summarize, the spacetime causal structure of the metrics (\ref{f}) is shown in Fig. \ref{pd}. For the case with odd power $n$, there are three roots for $f(r)=0$, denoted as $r_+$, $r_-$, $r'_+$, which correspond to three horizons. The region I is our asymptotically flat spacetime region and $r_+$ is the event horizon. Contrary to the Reissner-Nordström spacetime, the metric has a regular core in $r=0$ that forces us to extend the spacetime to $r<0$. However, it can not be extended beyond the region IV because the geodesics end at the singularity located in $r=-L$. For the case with even power $n$, $f(r)=0$ has two positive roots, denoted as $r_+$ and $r_-$ in Fig \ref{pd}. The region I and III, namely $-\infty < r < r_-$ and $r_+ < r< \infty$, are both asymptotically flat, while the region II interconnects two asymptotic regions of the Universe. The spacetime is geodetically complete for massless as well as massive particles.

%%%%%%%%%%%%%%%%%%%%%%%%%%%%%%%%%%%%%%%%%%%%%%%%%%%
\section{A geodetically incomplete black hole with Minkowski core}
\begin{figure}
\begin{center}
\includegraphics[height=5.6cm]{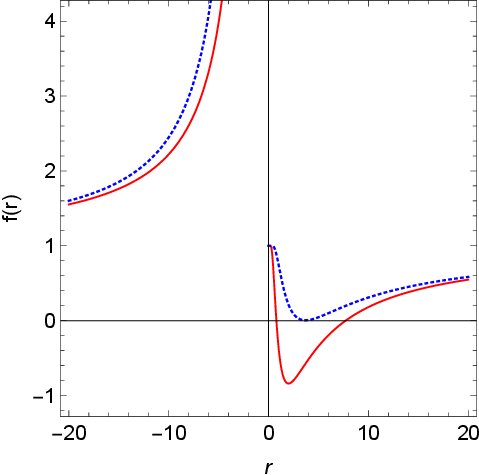}
\end{center}
 \caption{We here plot the function $f(r)$ for the case of $a<2M/{\rm e}$ (solid line), and $a=2M/{\rm e}$ (dashed line). In both cases $f(r)$ is not analytic in $r=4$. We fixed the mass to the value  $M=5$.}
    \label{visserf}
\end{figure}

In this section, we focus on the regular non-analytic black hole proposed by Culetu \cite{Culetu:2013fsa, Culetu:2014lca}, Simpson, and Visser in \cite{Simpson:2019mud}, and defined in terms of the following function, 
\be
\label{vissermetric}
f(r)= 1 - \frac{2M}{r}{\rm e}^{-a/r}\, .
\ee
In the $r \rightarrow 0^+$ limit, the function $f(r)$ reads $f(r)\sim 1- \mathcal{O}(r^2)$, and the black hole metric has an asymptotic Minkowski core where the curvature invariants vanish. However, the metric in $r=0$ is not continuous as it is easy to check computing the limit $r\rightarrow0^-$ of the function (\ref{vissermetric}) (see Fig. \ref{visserf}).

Let us now check if the metric is geodetically complete, namely if the proper time to reach the Minkowski core in $r=0$ is infinity. 
The geodesic equation of motion for radially falling massive particles reads:
\be
\dot{r}^2=e^2-1+\frac{2M}{r} {\rm e}^{-a/r}\, , 
\label{GeoViss}
\ee
%where $e$ is the conserved dimensionless energy. 
while the effective potential $V=f(r)$ has been plotted in Fig. \ref{visserf}. When the dimensionless constant $e^2<1$ the probe particles cannot reach $r=0^+$ and will bounce back. On the other hand for probe particles with $e^2=1$ we have $d\tau\propto r \, {\rm e}^{a/r}dr$, and the proper time to reach $r=0^+$ turns out to be infinite. Finally, for $e^2>1$, $d\tau/dr$ tends to the constant $e^2-1$ when $r\rightarrow0^+$. Therefore, the latter probe particles can reach the Minkowski core in finite proper time.

Contrary to the Hayward spacetime, any continuation of the geodesics is non-analytic. If the analyticity condition cannot be satisfied, there are many workable ways for the continuation of geodesics that can be chosen. As stated in the introduction, we here extend the coordinate $r$ to negative values and choose a simple extension that takes the same form of the metric (\ref{vissermetric}).
%\footnote{\color{red} As shown in \cite{Bambi:2013ufa}, we usually take a complexification of coordinate $r$ to obtain Kerr-like metrics. In this way, the form of the function (\ref{vissermetric}) is the same in the entire complex plane, except for the original point. It implies that the rotating Culetu-Simpson-Visser metric obtained by the Newman-Janis algorithm can be analytically extended to negative values of $r$. Therefore, the rotating black hole metric is the same for both positive and negative values of $r$. Considering the spherically symmetric black holes as the small spin limit of rotating black holes, we should take the same function (\ref{vissermetric}) as the metric of the negative real axis of $r$.}. 

Since the geodesic equation (\ref{GeoViss}) (see also Fig. \ref{visserf}) is ill-defined at $r=0$, the geodesics that end in $r=0$ can not be extended to negative values of $r$. Hence, it turns out that the metric (\ref{vissermetric}) is actually geodetically incomplete whether we assume the metric to be defined by the function (\ref{vissermetric}) also for negative values of $r$. 
\begin{figure}
\begin{center}
\includegraphics[height=5.6cm]{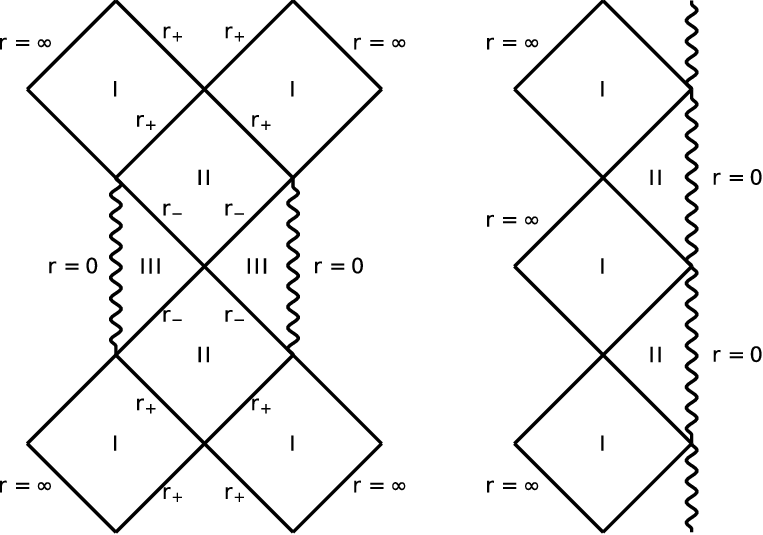}
\end{center}
\caption{The Penrose diagram for the maximal extension of the metric (\ref{vissermetric}).}
\label{pic5}
\end{figure}

Let us end this section by studying the spacetime causal structure of the metric (\ref{vissermetric}), which is closely related to the value of parameter $a$ present in (\ref{vissermetric}). For the case of $a < 2M/{\rm e}$ (here ${\rm e}$ is the Euler's number that should not be confused with the dimensionless energy $e$), the equation $f(r)=0$ have two solutions, usually denoted as $r_+$ and $r_-$. Therefore, the spacetime structure, in this case, is similar to Reissner-Nordström spacetime. For the case of $a=2M/{\rm e}$, the two horizons merge together, namely, $r_+=r_-=2M/{\rm e}$. Finally, for $a>2M/{\rm e}$, there is no horizon and the metric does not describe a black hole, but a kind of soliton. The Penrose diagram for the case $a<2M/{\rm e}$ and for the extreme case $a=2M/{\rm e}$ are given in Fig. \ref{pic5}.

%%%%%%%%%%%%%%%%%%%%%%%%%%%%%%%%%
\section{Geodetically complete black hole with Minkowski core}

In order to avoid the geodesic incompleteness of the metric (\ref{vissermetric}), we propose a slight modification to the Simpson-Visser's metric, namely \cite{Xiang:2013sza}
\be\label{improvedvisser}
f(r)= 1 - \frac{2M}{r} {\rm e}^{-a^2/r^2}\, .
\ee
The metric tends to the Schwarzschild metric for large values of the radial coordinate $r$ and has a Minkowski core for  $r\rightarrow0^+$. Likewise the Culetu-Simpson-Visser's metric, probe massive particles can reach $r=0$ in finite proper time, and we are again forced to make a continuation of the spacetime beyond $r=0$. However, contrary to the metric (\ref{vissermetric}), the improved metric (\ref{improvedvisser}) is continuous in $r=0$, and the extension of the spacetime to negative $r$ is doable.
\begin{figure}
\begin{center}
\includegraphics[height=5.5cm]{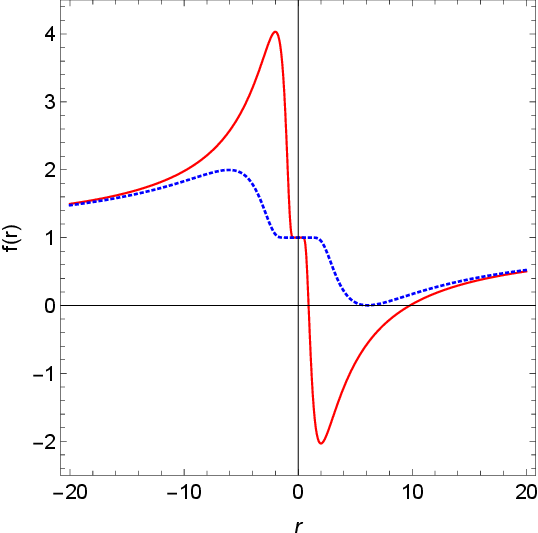}
\end{center}
 \caption{The function $f(r)$ for the case $a<\sqrt{2/ {\rm e}}M$ (solid line), and $a=\sqrt{2/ {\rm e}}M\approx 0.858M$ (dashed line). In both plots $M=5$.}
    \label{visserf2}
    \end{figure}

A plot of the function $f(r)$ in (\ref{improvedvisser}) is provided in Fig. \ref{visserf2}, which is similar to the generalized Hayward metric (\ref{f}) for even powers of the integer $n$. Moreover, the spacetime described by the metric (\ref{improvedvisser}) is geodetically complete because the proper time for massive particles to reach the edges at $r=\pm \infty$ of the spacetime is infinite.
\begin{figure}
\begin{center}
\includegraphics[height=5.5cm]{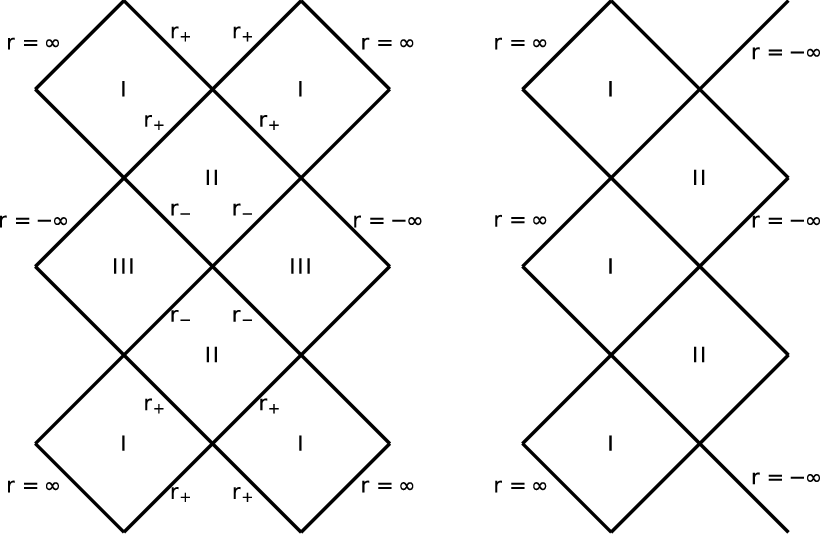}
\end{center}
 \caption{The Penrose diagram for the maximally extended metric (\ref{improvedvisser}).}
    \label{improvedvisserpd}
    \end{figure}

The spacetime structure of the metric (\ref{improvedvisser}) is given in Fig. \ref{improvedvisserpd}. For the case of $a^2<2M^2/{\rm e}$, there are two horizons, denoted by $r_+$ and $r_-$. The regions $r = \pm \infty$ represent the causal infinity of the region I and the region III respectively. The Minkowski core in $r=0$, which is located inside the region III, is regular. For the extreme case, namely $a^2=2M^2/{\rm e}$, the two horizons $r_+$ and $r_-$ coincide, and the spacetime consists of two asymptotically flat regions.

A more general class of geodetically complete modifications of the Culetu-Simpson-Visser spacetime looks like \cite{Xiang:2013sza}:
\be\label{improvedvisserG}
f(r)= 1 - \frac{2M}{r} {\rm e}^{- \frac{a^{2 n}}{r^{2 n}} }\, , \quad n \in \mathbb{N} \, , \quad n \geqslant 1.
\ee

The continuation of the metric (\ref{SchLINE}) with the function (\ref{improvedvisserG}) 
assumes the same form of $f(r)$ for positive as well as negative values of the radial coordinate $r$. However, if we assume $f(r)$ to be a $C^\infty$ smooth function the spacetime metric defined in terms of (\ref{improvedvisserG}) is not unique because the function $f(r)$ is non-analytic at $r=0$ since any derivative of $f(r)$ is zero in $r=0$. Therefore, as an even simpler choice, we could just extend the metric for $r\in (-\infty,0)$ to be exactly Minkowski.
Notice that the latter extension to Minkowski works also for the Culetu-Simpson-Visser's metric (\ref{vissermetric}), which in this way turns out to be geodetically complete. 
In other words, the simplest extension of the Culetu-Simpson-Visser metric (consisting in taking the function (\ref{vissermetric}) also for $r <0$) is not geodetically complete, but its extension to Minkowski or to other smooth functions will be consistent with the geodesic completion. 

Finally, we make a comment about the uniqueness issue. Indeed, multiple (in principle infinity) extensions of the metric imply multiple evolutions for point-like particles beyond $r=0$, which boils down to the loss of predictivity. In other words, $r=0$ turns out to be a bifurcation or multi-furcation point once the metric is fixed for $r>0$ because of the ambiguous extension to negative values of the radial coordinate.  
%However, this is not a problem related to the geodetic principle, but to the uniqueness of the metric. 
Such a multi-furcation problem is tantamount of the geodesic incompletion because in both cases manifest a loss of predictability. Therefore, we here face two options: (i) we reject non analytic spacetimes, or (ii) we select a class of non analytic spacetime whose all extensions are geodeticaly complete.  
In this paper, we assumed (ii).

%In this respect will be crucial to have a theory of which one of the regular spacetimes proposed in this section will be an exact solution. 
%{\color{red} However, in absence of such theory, in order to take seriously non analytic metrics as a resolution of the singularity problem we require all the possible extensions to be geodetically complete. Therefore, the antipodal and the negative $r$ extension must be both complete.}

%As a side effect, we found that when one relaxes analyticity uniqueness is lost. 

%, the metric is also $C^\infty$ continuous in the entire spacetime.

%\vspace{10cm}

%%%%%%%%%%%%%%%%%%%%%%%%%%%%%%%%% Conclusion
\section{Summary and Discussion}

In this paper, we focused on the geodesic completeness of some popular singularity-free black holes whose regularity is based on curvature invariants. 
We pointed out that in order to ensure the analyticity of the geodesic equations, geodetics reaching $r=0$ have to be extended to negative values of $r$ if the metric exhibits analyticity.
 %but not even functions of $r$. 
 As a typical example, the Hayward spacetime should be %accordingly
  extended to negative values of $r$.  Hence, we showed that the spacetime is geodetically incomplete because the Hayward metric (\ref{heywardf}) is singular at the point $r= - L$. 
  
In the case of non-analytic regular black holes, we cannot require the analyticity of geodesic equations and the continuation of geodesics is not unique. Hence, such nonanalytic metrics can not in general solve the singularity issue. Indeed, {\em the ambiguity in the extension of the metric beyond $r=0$ is tantamount of geodesic incompleteness}. In order, to avoid such issue we select as physical nonanalytic spacetimes the ones whose all extensions are geodetically complete. 
    In such respect, we focused on the simplest extension of non-analytic metrics, namely to negative values of $r$ in order to keep the metric form-invariant. Hence, we showed that the simplest extension of the non-analytic metric with a Minkowski core proposed by Culetu, Simpson, and Visser (\ref{vissermetric}) is not geodetically complete. Therefore, such spacetimes are actually pathological although $r=0$ is a smooth core. To address such issue, we proposed a generalization of the metrics of above and showed that the modified metrics are geodetically complete. Although the proposed complete spacetimes are a simple extension of Hayward and Culetu-Simpson-Visser's metrics, we think they need an analysis of the thermodynamic properties that we have not addressed in this paper. In particular, the gravitational collapse should be rethought in light of the causal structure derived in this article. 

%{\color{red}Generally speaking, for spherically symmetric regular black holes that metrics are not even functions of $r$, it is necessary to check the regularity of metrics for any $r\in [-\infty, +\infty]$}.

In particular, one can show the geodesic completeness of other popular regular black holes. For example, the {\em Bardeen's black hole} \cite{Bardeen}, the {\em Nicolini-Samilagic-Spallucci's black hole} \cite{Nicolini:2005vd, Modesto:2010uh, Burzilla:2020utr, Giacchini:2018wlf}, the {\em Lee-Wick black hole} \cite{Bambi:2016wmo}, and the {\em loop black hole} \cite{Modesto:2008im, Modesto:2009ve} and other singularity-free spacetimes found in conformal gravity are geodetically complete \cite{Bambi:2016wdn}. 
As a further feature and contrary to the other regular black holes, conformal black holes, which are a prediction of local \cite{Modesto:2015ozb,Modesto:2016ofr} and nonlocal \cite{Modesto:2011kw,Modesto:2017sdr,Modesto:2014lga,Modesto:2021okr,Modesto:2021ief} higher derivative theories, do not show any Cauchy horizon (if not present in the solution of Einstein's theory), usually unstable \cite{Poisson:1990eh} in the Einstein's theory of gravity.

%Similarly, we can check the geodesic completeness of other popular regular black hole models. For example, the Bardeen's black hole\cite{Bardeen}, Nicolini's black hole\cite{Modesto:2010uh}, and Modesto's black hole in loop quantum gravity\cite{Modesto:2009ve} are geodetically complete. For these complete spacetime, including the generalized Hayward metric with even power $n$ and the modified non-analytic black hole we have discussed in this paper, there are always multi-horizon structures, which make these complete black holes unstable because of the mass inflation. So stability is also a considerable issue of complete black hole models.

%\vspace{1cm}

%%%%%%%%%%%%%%%%%%%
\section*{ACKNOWLEDGMENTS}
We are grateful to Breno L. Giacchini, Daniel Terno, and Cosimo Bambi for the useful comments and questions. This work has been supported by the Basic Research Program of the Science, Technology, and Innovation Commission of Shenzhen Municipality (grant no. JCYJ20180302174206969). 

%%%%%%%%%%%%%%%%%%%%%%%%%%

\appendix

\section{Analytic continuation of geodesics}

In this section, we show that we are forced to extend some spacetimes to negative values of $r$ whether the analytic continuation of geodesics is required. In particular, for the Hayward spacetime, the equation of motion of ingoing radial null geodesic is given by
\be
-f(r)\dot{t}^2+f^{-1}(r)\dot{r}^2=0\, ,\quad  f(r)=1-\frac{2Mr^2}{r^3+L^3}\, ,
\label{pippo}
\ee
where the dot represents the derivative with respect to the affine parameter $\lambda$. Replacing the conserved quantity $e \equiv -g_{tt}\dot{t}$ in (\ref{pippo}), we get
\be\label{nulleom}
\dot{r}^2=e^2\, ,\qquad  \dot{t}^2= e^2 \left( 1- \frac{2Mr^2}{r^3+L^3} \right)^{-2}\, .
\label{puppa}
\ee
Since $\dot{r}$ is constant, we have $\Delta\lambda\propto \Delta r$ and the affine parameter to reach $r=0$ is finite. 

\begin{figure}[ht]
\begin{center}
\includegraphics[height=3.2cm]{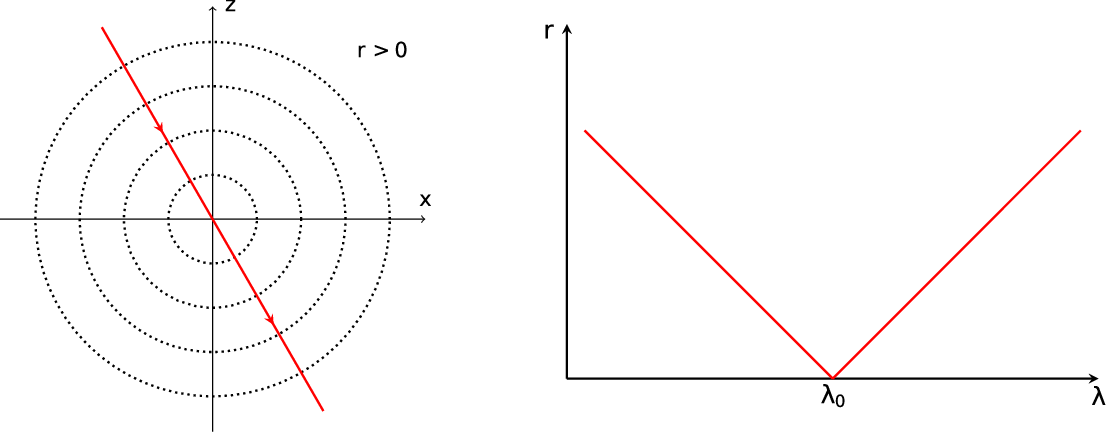}
\end{center}
 \caption{Panel on the left: the simple antipodal continuation of radial geodesics. Panel on the right: the function $r(\lambda)$ for the null geodesics described by the equation (\ref{nulleom}).}
    \label{antipodal}
\end{figure}

\begin{figure}[ht]
\begin{center}
\includegraphics[height=3.5cm]{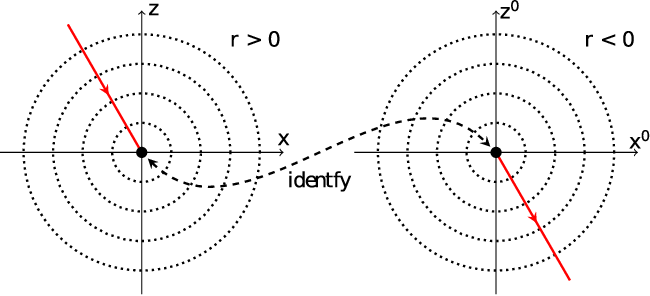}
\end{center}
 \caption{The continuation of geodesics to negative values of $r$.}
  \label{analytic}
\end{figure}

Let us now consider the analytic continuation of the geodesic. There are two possible continuations we can consider: (i) the coordinate $r$ is always positive then the geodesics should be continued to the antipodal direction (see Fig.\ref{antipodal}), namely
\be
\theta\rightarrow \pi-\theta\, , \ \phi\rightarrow \phi+\pi\, ;
\ee
(ii) the geodesics should be continued to negative values of $r$ (see Fig.\ref{analytic}). 
The antipodal continuation of geodesics is the simplest one, but it is not analytic. In order to show this issue, we can verify the analyticity of the parametric equation $t(\lambda)$ in (\ref{puppa}). %It is required that, under 
In Cartesian coordinates, the parametric curves $\{t(\lambda), \, x(\lambda), \, y(\lambda), \, z(\lambda) \}$ of should be analytic functions. From the geodesic equation (\ref{nulleom}), $\dot{t}$ can be expanded around $r=0$ as
\be\label{expanddott}
\dot{t} (r) = 1+ \frac{2M}{L^3}r^2 + \frac{4M^2}{L^6} r^4 -\frac{2M}{L^6} r^5 + \mathcal{O}(r^5)    \, ,
\ee 
where, without loss of generality, we fixed $e=1$ and the condition $\dot{t}>0$ is used since $t$ is a timelike coordinate near $r=0$. The $r^5$ term in the Taylor's expansion (\ref{expanddott}) implies that the fifth-order derivative of $\dot{t}(\lambda)$ is non-zero and changes sign in $r=0$. Indeed, 
\be\label{5dot}
\left(\frac{d}{d\lambda}\right)^5 \dot{t}(\lambda) = \dot{r}^5 \left(\frac{d}{dr}\right)^5 \dot{t}(r)\, .
\ee
As shown in Fig.\ref{antipodal}, when a test particle crosses the origin $r=0$, the function $\dot{r}$ changes the sign instantaneously, namely $\dot{r}= -1 \rightarrow +1$ for $\lambda = \lambda_0$. Therefore, the left and right limits of formula (\ref{5dot}) are different at $r=0$, namely
\begin{align}
\lim_{\lambda\rightarrow \lambda_0} \left(\frac{d}{d\lambda}\right)^5 \dot{t}(\lambda) = \frac{240M}{L^6}\, &,\ {\rm when}\ \lambda<\lambda_0\, , \nonumber \\
\lim_{\lambda\rightarrow \lambda_0} \left(\frac{d}{d\lambda}\right)^5 \dot{t}(\lambda) = -\frac{240M}{L^6}\, &,\ {\rm when}\ \lambda>\lambda_0\, .
\end{align}
The discontinuity of $(d/d\lambda)^6 t(\lambda)$ at $r=0$ breaks the analyticity of the geodesic equation. 

For the continuation shown in Fig.\ref{analytic}, since $\dot{r}$ does not change the sign, the analyticity of function $t(\lambda)$ is preserved. Hence, the continuation to negative values of $r$ is the only choice we have to make for the case of the Hayward spacetime consistently with the analyticity of geodesics.

In general, %for the continuation problem, generic 
spherically symmetric metrics with well-defined Taylor expansion near $r=0$ %of the metric function $f(r)$ 
can be divided into two classes:
\begin{itemize}
\item[(1).] The metric is not symmetric under $r\leftrightarrow -r$. 
This class includes the Hayward spacetime\cite{Hayward:2005gi} and its generalizations. The terms with odd powers of $r$ are present in the Taylor expansion of the metrics thus the antipodal continuation is not analytic. Therefore, this kind of spacetimes should be continued beyond $r=0$ to negative values of $r$.

\item[(2).] The metric is symmetric under $r\leftrightarrow -r$, such as the Minkowski spacetime and the Bardeen black hole\cite{Bardeen}. Since all the powers of terms in the Taylor series of metrics are even, the simple antipodal continuation is analytic. So it is not necessary to extend the coordinate $r$ to negative values for such spacetimes. Moreover, predictably, there is no novel result that motivates us to extend such spacetimes to negative values of $r$ because the extension is an isometry.
\end{itemize} 

Finally, it deserves to be noticed that the above conclusion is similar to the main result of Ref.\cite{Giacchini:2021pmr}, which shows that terms with odd powers of $r$ in the Taylor series of metric around $r=0$ make higher-derivative curvature invariants to diverge. Indeed, using Cartesian coordinates near $r=0$, it is easy to verify that the first kind of metrics we summarized above are not smooth, which causes the divergence of higher-derivative curvature invariants and the non-analyticity of geodesic equations. Moreover, in \cite{Giacchini:2021pmr} it was proved that this kind of metrics might not contribute to the Lorentzian path integral if the classical gravitational action contains higher-derivative curvature invariants. In this paper, we mainly focus on the analyticity of geodesic equations and we stress that spacetimes with the first kind of metrics should be extended to negative values of $r$. 


\begin{thebibliography}{99}

%%%%%% regular black hole models

\bibitem{Bardeen}
J.~M.~Bardeen, in 
``Conference Proceedings of GR5 (Tbilisi, USSR, 1968)", p. \textbf{174}.
%M.~J.~Bardeen, in 
%``Conference Proceedings of GR5'' (Tbilisi, USSR, 1968), pag. 174. 
%Abstracts of the 5th International Conference on Gravitation and the Theory of Relativity (Tbilisi, USSR),''
%Tbilisi University Press. \textbf{174} (1968).

%\cite{Hayward:2005gi}. Heyward
\bibitem{Hayward:2005gi}
S.~A.~Hayward,
``Formation and evaporation of regular black holes,''
Phys. Rev. Lett. \textbf{96}, 031103 (2006).
%doi:10.1103/PhysRevLett.96.031103
[arXiv:gr-qc/0506126 [gr-qc]].
%671 citations counted in INSPIRE as of 27 May 2022


\bibitem{nature}
J.~D.~Barrow and F.~J.~Tipler,
``Action principles in nature,''
Nature \textbf{331}, 31-34 (1998).

%\cite{Giacchini:2021pmr}
\bibitem{Giacchini:2021pmr}
B.~L.~Giacchini, T.~d.~Netto and L.~Modesto,
``Action principle selection of regular black holes,''
Phys. Rev. D \textbf{104}, no.8, 084072 (2021)
%doi:10.1103/PhysRevD.104.084072
[arXiv:2105.00300 [gr-qc]].



%\cite{Culetu:2013fsa}
\bibitem{Culetu:2013fsa}
H.~Culetu,
``On a regular modified Schwarzschild spacetime,''
arXiv:1305.5964 [gr-qc]].

%\cite{Culetu:2014lca}
\bibitem{Culetu:2014lca}
H.~Culetu,
``On a regular charged black hole with a nonlinear electric source,''
Int. J. Theor. Phys. \textbf{54}, no.8, 2855-2863 (2015)
%doi:10.1007/s10773-015-2521-6
[arXiv:1408.3334 [gr-qc]].

%\cite{Xiang:2013sza}
\bibitem{Xiang:2013sza}
L.~Xiang, Y.~Ling and Y.~G.~Shen,
``Singularities and the Finale of Black Hole Evaporation,''
Int. J. Mod. Phys. D \textbf{22}, 1342016 (2013)
%doi:10.1142/S0218271813420169
[arXiv:1305.3851 [gr-qc]].



%\cite{Simpson:2019mud}. Visser
\bibitem{Simpson:2019mud} 
A.~Simpson and M.~Visser,
``Regular black holes with asymptotically Minkowski cores,''
Universe \textbf{6}, no.1, 8 (2019).
%doi:10.3390/universe6010008
%[arXiv:1911.01020 [gr-qc]].
%48 citations counted in INSPIRE as of 27 May 2022

%\cite{Bonanno:2000ep}
\bibitem{Bonanno:2000ep}   %Asymptotically safe gravity
A.~Bonanno and M.~Reuter,
``Renormalization group improved black hole space-times,''
Phys. Rev. D \textbf{62}, 043008 (2000); 
%\cite{Bonanno:2006eu}
%\bibitem{Bonanno:2006eu}
A.~Bonanno and M.~Reuter,
``Spacetime structure of an evaporating black hole in quantum gravity,''
Phys. Rev. D \textbf{73}, 083005 (2006).
%doi:10.1103/PhysRevD.73.083005
%[arXiv:hep-th/0602159 [hep-th]].
%201 citations counted in INSPIRE as of 06 Jul 2022
%doi:10.1103/PhysRevD.62.043008
%[arXiv:hep-th/0002196 [hep-th]].
%407 citations counted in INSPIRE as of 06 Jul 2022

\bibitem{Nicolini:2005vd}   
P.~Nicolini, A.~Smailagic and E.~Spallucci,
``Noncommutative geometry inspired Schwarzschild black hole,''
Phys. Lett. B \textbf{632}, 547 (2006),
gr-qc/0510112.


%\cite{Modesto:2010uh}
\bibitem{Modesto:2010uh} %% Nicolini
L.~Modesto, J.~W.~Moffat and P.~Nicolini,
``Black holes in an ultraviolet complete quantum gravity,''
Phys. Lett. B \textbf{695}, 397-400 (2011)
%doi:10.1016/j.physletb.2010.11.046
[arXiv:1010.0680 [gr-qc]].
%139 citations counted in INSPIRE as of 27 May 2022


%\cite{Burzilla:2020utr}
\bibitem{Burzilla:2020utr}
N.~Burzill\`a, B.~L.~Giacchini, T.~d.~Netto and L.~Modesto,
``Higher-order regularity in local and nonlocal quantum gravity,''
Eur. Phys. J. C \textbf{81}, no.5, 462 (2021)
%doi:10.1140/epjc/s10052-021-09238-x
[arXiv:2012.11829 [gr-qc]].
%12 citations counted in INSPIRE as of 26 Jul 2022


%\cite{Giacchini:2018wlf}
\bibitem{Giacchini:2018wlf}
B.~L.~Giacchini and T.~de Paula Netto,
``Effective delta sources and regularity in higher-derivative and ghost-free gravity,''
JCAP \textbf{07}, 013 (2019)
%doi:10.1088/1475-7516/2019/07/013
[arXiv:1809.05907 [gr-qc]].
%42 citations counted in INSPIRE as of 26 Jul 2022




%\cite{Modesto:2008im}   %LQG
\bibitem{Modesto:2008im}
L.~Modesto,
``Semiclassical loop quantum black hole,''
Int. J. Theor. Phys. \textbf{49}, 1649-1683 (2010)
%doi:10.1007/s10773-010-0346-x
[arXiv:0811.2196 [gr-qc]].
%117 citations counted in INSPIRE as of 06 Jul 2022

%\cite{Modesto:2009ve}
\bibitem{Modesto:2009ve}  %LQG
L.~Modesto and I.~Premont-Schwarz,
``Self-dual Black Holes in LQG: Theory and Phenomenology,''
Phys. Rev. D \textbf{80}, 064041 (2009)
%doi:10.1103/PhysRevD.80.064041
[arXiv:0905.3170 [hep-th]].
%69 citations counted in INSPIRE as of 27 May 2022

%\cite{Bambi:2016wmo}  %lee-wick BH
\bibitem{Bambi:2016wmo}
C.~Bambi, L.~Modesto and Y.~Wang,
%``Lee\textendash{}Wick black holes,''
Phys. Lett. B \textbf{764}, 306-309 (2017)
doi:10.1016/j.physletb.2016.11.060
[arXiv:1611.03650 [gr-qc]].
%14 citations counted in INSPIRE as of 05 Jul 2022


%\cite{Bambi:2016wdn}. % conformal BH
\bibitem{Bambi:2016wdn}
C.~Bambi, L.~Modesto and L.~Rachwa\l{},
``Spacetime completeness of non-singular black holes in conformal gravity,''
JCAP \textbf{05}, 003 (2017)
%doi:10.1088/1475-7516/2017/05/003
[arXiv:1611.00865 [gr-qc]].
%73 citations counted in INSPIRE as of 01 Jul 2022






%LOCAL


%\cite{Modesto:2015ozb}
\bibitem{Modesto:2015ozb}
L.~Modesto and I.~L.~Shapiro,
``Superrenormalizable quantum gravity with complex ghosts,''
Phys. Lett. B \textbf{755}, 279-284 (2016)
%doi:10.1016/j.physletb.2016.02.021
[arXiv:1512.07600 [hep-th]].
%124 citations counted in INSPIRE as of 03 Aug 2022

%\cite{Modesto:2016ofr}
\bibitem{Modesto:2016ofr}
L.~Modesto,
``Super-renormalizable or finite Lee\textendash{}Wick quantum gravity,''
Nucl. Phys. B \textbf{909}, 584-606 (2016)
%doi:10.1016/j.nuclphysb.2016.06.004
[arXiv:1602.02421 [hep-th]].
%93 citations counted in INSPIRE as of 03 Aug 2022




%NONLOCAL

%\cite{Modesto:2011kw}
\bibitem{Modesto:2011kw}
L.~Modesto,
``Super-renormalizable Quantum Gravity,''
Phys. Rev. D \textbf{86}, 044005 (2012)
%doi:10.1103/PhysRevD.86.044005
[arXiv:1107.2403 [hep-th]].
%442 citations counted in INSPIRE as of 03 Aug 2022



%\cite{Modesto:2017sdr}
\bibitem{Modesto:2017sdr}
L.~Modesto and L.~Rachwa\l{},
``Nonlocal quantum gravity: A review,''
Int. J. Mod. Phys. D \textbf{26}, no.11, 1730020 (2017)
%doi:10.1142/S0218271817300208
%93 citations counted in INSPIRE as of 03 Aug 2022


%\cite{Modesto:2014lga}
\bibitem{Modesto:2014lga}
L.~Modesto and L.~Rachwal,
``Super-renormalizable and finite gravitational theories,''
Nucl. Phys. B \textbf{889}, 228-248 (2014)
%doi:10.1016/j.nuclphysb.2014.10.015
[arXiv:1407.8036 [hep-th]].
%182 citations counted in INSPIRE as of 03 Aug 2022




%\cite{Modesto:2021okr}
\bibitem{Modesto:2021okr}
L.~Modesto,
``The Higgs mechanism in nonlocal field theory,''
JHEP \textbf{06}, 049 (2021)
%doi:10.1007/JHEP06(2021)049
[arXiv:2103.05536 [hep-th]].
%9 citations counted in INSPIRE as of 03 Aug 2022


%\cite{Modesto:2021ief}
\bibitem{Modesto:2021ief}
L.~Modesto,
``Nonlocal Spacetime-Matter,''
[arXiv:2103.04936 [gr-qc]].
%10 citations counted in INSPIRE as of 03 Aug 2022




%\cite{Poisson:1990eh}. % Instability of Cauchy horizon
\bibitem{Poisson:1990eh}
E.~Poisson and W.~Israel,
``Internal structure of black holes,''
Phys. Rev. D \textbf{41}, 1796-1809 (1990); 
%doi:10.1103/PhysRevD.41.1796
%578 citations counted in INSPIRE as of 05 Jul 2022
%\cite{Ori:1991zz}
%\bibitem{Ori:1991zz}
A.~Ori,
``Inner structure of a charged black hole: An exact mass-inflation solution,''
Phys. Rev. Lett. \textbf{67}, 789-792 (1991).
%doi:10.1103/PhysRevLett.67.789
%250 citations counted in INSPIRE as of 05 Jul 2022

\end{thebibliography}
\end{document}